\documentclass[reprint,amsmath,amssymb,aps,showkeys,superscriptaddress]{revtex4-2}

\usepackage{graphicx}																									
\usepackage{titlesec}
\usepackage{amsmath}	
\usepackage{enumitem}	
\usepackage{soul}

\usepackage{mathabx}												 
\usepackage{fancyhdr}	
\usepackage{physics}
\usepackage{subcaption}
\usepackage[usenames,dvipsnames]{color}% color text
\usepackage[normalem]{ulem} % przekreslenia: \sout{Hello World}
\allowdisplaybreaks

\makeatletter
\def\hlinewd#1{%
  \noalign{\ifnum0=`}\fi\hrule \@height #1 \futurelet
   \reserved@a\@xhline}
\makeatother

\begin{document}

\title{Description of molecular nanomagnets by the multi-orbital Hubbard model \\ with correlated hopping}

\author{J. Matysiak}
\email[e-mail: ]{j.matysiak@intibs.pl}
\affiliation{Institute of Low Temperature and Structure Research, Polish Academy of Science, ul. Ok\'{o}lna 2, 50-422 Wroc\l{}aw, Poland}

\author{R. Lema\'{n}ski}
\email[Corresponding author; e-mail: ]{r.lemanski@intibs.pl}
\affiliation{Institute of Low Temperature and Structure Research, Polish Academy of Science, ul. Ok\'{o}lna 2, 50-422 Wroc\l{}aw, Poland}

\date{\today}

\begin{abstract}
We present a microscopic description of molecular magnets by the multi-orbital Hubbard model, which includes the correlated hopping term, i.e. the dependence of electron hopping amplitude between orbitals on the degree of their occupancy. 
In the limit of large Coulomb on-site interaction, we derived the spin Hamiltonian using the perturbation theory. 
The magnetic coupling constant between two ions we determined in two different ways: a)from the expression obtained in the perturbation calculus and b)from the analysis of distances between the lowest levels of the energy spectrum obtained by the diagonalization of the multi-orbital Hubbard model.
The procedure we use can be applied to various nanomagnets, but the final calculations we performed for the molecular ring $Cr_8$. We showed that the correlated hopping can reduce the antiferromagnetic exchange between ions, what is essential for a proper description of $Cr_8$.
\end{abstract}



\maketitle

\section{Introduction}
Molecular nanomagnets (MNMs) are molecules containing a core of a finite number of d or f ions whose spins are magnetically coupled. These systems are becoming more and more popular due to their potential use in constructions of, among others, quantum computers and high-performance magnetic memories \cite{Meier2003,Troiani2005,Hao2015}. 

In the theoretical studies of MNMs, a key issue is the calculation of the coupling parameters between magnetic ions. This is quite a difficult task due to the presence of unfilled d or f shells and dynamical electron correlations, which can be described by the Hubbard-type models. However, it turns out that typical spectra of low-energy excitations of MNMs determined experimentally correspond (approximately) to the excitation spectra obtained for models of localized spins with dominant isotropic Heisenberg-type coupling \cite{Furrer2013}. For this reason, the efforts of many researchers have been directed towards determining the exchange coupling constants in a system-appropriate version of the Heisenberg model. 
The calculations are usually performed with methods based on the density functional theory (DFT), which enable determination from first principles of various microscopic parameters for specific materials \cite{Furrer2013,Kortus2001,Gatteschi2006,Gao2015}. In fact, there are many variants of the DFT-based method and the literature on this subject is extensive (see for example Refs.
\cite{Katsnelson2000,Kortus2001,Boukhalov2002,Boukhalov2004,Mazurenko2014,Hanebaum2015,Schnack2019}).
However, in some cases the DFT-based methods do not give satisfactory results. 
For example, in some \emph{chromium based rings} the excitations energies derived from the DFT calculations  substantially outweigh values measured in experiments \cite{Brzostowski2013,Brzostowski2014}.  

In this case, the problem with finding the correct value of the exchange parameter seems to arise from the difficulty in determining the eigenstates of the Heisenberg model. Indeed, the presence of spin-flipping terms in the Heisenberg model causes that the fixed spin configurations (except the ferromagnetic one) are not eigenstates of the Heisenberg Hamiltonian, therefore they energies are not fixed. Consequently, calculations in which the exchange parameter is determined on the basis of the energy difference of states with fixed spin configurations (e.g. ferromagnetic and antiferromagnetic of the Neel's type) do not lead to the correct results.
Although quite recently it was reported that the magnetic couplings calculated using the symmetry-broken version of DFT are close to those deduced from the experimental data \cite{WeissmanJCTC2019}.

In general, the problem with obtaining correct magnetic coupling values by DFT-based methods stems from the lack of consideration of electron dynamics and the resulting electron correlations in these methods. A possible remedy to this problem was proposed, e.g., in  \cite{Katsnelson2000}, where LDA++ method was presented, which "deals with the thermodynamic potential as a functional of Green function rather than electron density". An extension of this approach referred to as the DFT + DMFT method is presented in \cite{Kvashnin2015}.

In fact, the problem of electron dynamics and correlation is one of the central issues of condensed matter theory and we will not discuss it further here. For more information on the advantages and limitations of different variants of the DFT method used in the calculation of the exchange constants of different MNMs, see e.g. Ref. \cite{Chiesa2013,Chiesa2016}.
It these papers it was also proposed to describe MNMs using the multi-orbital HM model combined with DFT calculations \cite{Chiesa2013,Chiesa2016}. In their approach, which they named the DFT + MB method (MB stands for many-body), they first determined the microscopic parameters of the multi-orbital Hubbard model (HM), and only then did they calculate the exchange constant using the second-order perturbation theory. 

The DFT + MB method seems to be universal because, starting from the multi-orbital HM, it takes into account the most important microscopic processes affecting the value of the exchange constant in the effective spin model, including the dynamical electron correlations \cite{KvashninPRB2015}. The correctness of this method seems to be confirmed by the results of the calculation of the exchange constants for several MNMs, which are given in \cite{Chiesa2013}. 

For example, for $Cr_8$ ring, the exchange constant $\Gamma = 1.65 meV$ appeared to be substantially lower than $3.5meV$ obtained from previous DFT calculations.
However, this value is still clearly greater than predicted from the experiments $\Gamma = 1.46 meV$, and also greater than that obtained from the new DFT calculations reported in \cite{WeissmanJCTC2019}.

To explain this discrepancy, here we extend the description of the system by including to the Hamiltonian the \emph{correlated hopping} (CH) term \cite{SchillerPRB1999, LuzPRB1996} (also known as \emph{correlated hybridization} \cite{Hubsch2006}), in which the amplitudes of the electron hopping between the orbitals depend on their occupancies.
More specifically, higher occupancies correspond to lower amplitudes. It results from the fact, that the hopping of an electron between two orbitals is hindered by the presence of another electron with opposite spin at the initial or final orbital. Obviously, it leads to a reduction of the hopping amplitude between orbitals if the sum of their occupancies is bigger than one, and consequently to a reduction in the antiferromagnetic exchange constant. 
In fact, CH has already been considered in the earlier works \cite{Vonsovsky}.

In real materials CH is always present, but its effect on physical properties of a given system is not always known, because it is rarely taken into account in theoretical descriptions. 
This is due to the fact, that the magnitude of the part of Coulomb interaction associated with CH is usually much smaller than the main part, which corresponds to the interaction between electrons located on the same orbital with opposite spins. At the same time, however, many authors admit that in spite of a relatively small strength it may play an essential role in the correct description of some systems \cite{CampbellPRB1988,CampbellProc1989,CampbellPRB1990,AppelPRB1993,Hirsch1994,LuzPRB1996,SchillerPRB1999,Hubsch2006,Vonsovsky,Galvan2012}. 
For example, in Ref. \cite{SchillerPRB1999} it is suggested that CH  ''may be essential for the complete understanding of the metal-insulator transition in $V_{2-y}O_3$ and $Ca_{1-x}Sr_xVO_3$'' while in Ref. \cite{Hubsch2006} it is demonstrated that CH ''may possibly provide a means to significant spin polarization of currents through transition-metal based molecular transistors in modest magnetic fields'' and that matrix elements resulting from CH ''could significantly change the parameters in effective single-band models for transition-metal oxides''.

In the case of 3\emph{d}-electrons in transition metals, Hubbard has already estimated the different contributions of the Coulomb interaction \cite{Hubbard1963}. Then also other authors estimated these parameters for various materials \cite{CampbellPRB1990,AppelPRB1993,Hirsch1994,SchillerPRB1999,Hubsch2006}. These estimates suggest that the correlated and direct hopping amplitudes have comparable magnitude, but typically the ratio of the former to the latter is less than 0.4. 

In this contribution, we investigate the effect of  CH on the value of  magnetic coupling constant between $Cr$ ions in the molecular ring $Cr_8$. Here we focused on $ Cr_8 $ because for this system we found data relevant to our calculations, but we suppose our considerations can be applied to other molecular nanomagnets as well. 

Taking an advantage of the fact that the hopping amplitudes are much smaller than the on-site Coulomb couplings, we first constructed the effective spin Hamiltonian using the perturbation theory. In this way we obtained the Heisenberg model with the exchange constant equal to $\Gamma$. The perturbative calculus to the multi-orbital HM was also applied in Ref. \cite{Chiesa2013}, but CH was not taken into account when constructing the effective Hamiltonian. Then we diagonalized the Hamiltonian (\ref{HamHubbard}) for the system of two $Cr$ ions and based on the analysis of the obtained exact energy spectrum we deduced another value $\Gamma^*$ for the exchange constant, which turned out to be slightly greater than $\Gamma$.  

The perturbative calculus for the multi-orbital HM was also applied in Ref. \cite{Chiesa2013}, but CH was not considered there in constructing the effective Hamiltonian and was also not included in our earlier comparative analysis of the multi-orbital HM and the Heisenberg model \cite{Matysiak}.

The remainder of the paper is organized as follows. 
In the next Sec. II we present the model, in Sec. III the exact solution for a single ion, in Sec. IV our main results regarding the interaction between two ions. The last Sec. V contains our brief summary and conclusions.

\section{Multi-orbital HM with correlated hopping }

The multi-orbital HM with CH $H_{HMcor}$ that we use here has the following form:

\begin{align*}
&H_{HMcor}=H_0+H_1 \stepcounter{equation}\tag{\theequation}
\label{HamHubbard}
\end{align*}
with the single-ion part\\
\begin{align*}
&H_0=U\sum\limits_{im} n_{im \downarrow}n_{im \uparrow}+ \\
&\frac{1}{2} \sum\limits_{i,m \neq m',\sigma} \left[ U'n_{im\sigma}n_{im'\bar{\sigma}}+U''n_{im\sigma}n_{im'\sigma} \right]+\\
& \frac{1}{2} \sum\limits_{i,m \neq m',\sigma} \left[ 
J c_{im\sigma}^\dagger c_{im'\bar{\sigma}}^\dagger c_{im\bar{\sigma}} c_{im'\sigma}+ \right.   
 \left. J c_{im\sigma}^\dagger c_{im\bar{\sigma}}^\dagger c_{im'\bar{\sigma}} c_{im'\sigma} \right]
\end{align*}
and the intersite hopping term\\
\begin{align*}
&H_1=\sum\limits_{\substack{i \neq j \\
m,m',\sigma}} t^{ij}_{mm'}[1-a(n_{im}+n_{jm'}-1)]c_{im\sigma}^\dagger c_{jm'\sigma}.
\end{align*}

In the above formulas $i$ and $j$ denote nearest-neighbor sites, $m,m'$ label orbitals and $\sigma,\bar{\sigma}$ label spins of electrons ($\bar{\sigma}=-\sigma$), 
$c^{\dag}_{im\sigma}$ ($c_{im\sigma}$) denotes the creation (annihilation) operator of an electron, 
$n_{im\sigma} = c^{\dag}_{im\sigma} c_{im\sigma}$ is the occupation number
and $n_{im}=n_{im\downarrow}+n_{im\uparrow}$.
$U$, $U'$ and $U''$ describe the Coulomb type on-site interactions between two electrons: $U$ - on the same orbital and $U'$($U''$) - on different orbitals with opposite (parallel) spins, respectively. $J$ represents the on-site exchange coupling resulting from the first Hund's rule. Here we adopt the following relations between parameters of the model: $U'=U-2J$ and $U''=U-3J$. They result from the requirements of rotational symmetry \cite{Fresard1997}.
The parameter $t^{ij}_{mm'}$ is the hopping amplitude ($i\neq j$) from orbital $m'$ at site $j$
to orbital $m$ at site $i$ or the energy $\varepsilon^{i}_{m}\equiv t^{ii}_{mm}$ of orbital $m$ at site $i$
($i=j$ and $m=m'$) and $a$ is the correlated hopping parameter.

The parameter $a$ in (\ref{HamHubbard}) is a measure of  reduction of the electron hopping amplitude between two orbitals when the sum of their occupancies exceeds one.
In real systems the degree of this reduction may be different for different pairs of orbitals. Consequently, the parameter $a$ may acquire appropriate indexes and be replaced by the table of parameters $a^{ii '}_{mm'}$.
However, since by now these quantities are not known from independent calculations, to avoid introducing many new parameters $a^{ii '}_{mm'}$, we assume here that for all relevant pairs of orbitals they have the same value $a^{ii ' }_{mm '}=a$. 
In fact, in all the studies that included CH we noticed so far \cite{CampbellPRB1988,CampbellProc1989,CampbellPRB1990,AppelPRB1993,Hirsch1994,LuzPRB1996,SchillerPRB1999,Hubsch2006,Galvan2012}, only systems of identical ions with one orbital per ion were considered, thus we have never noticed in the scientific literature the parameter $a$ with indexes.
But of course the value of $a$ depends on the material we are dealing with and, as we already mentioned before, the most common estimates were $0<a<0.4$ \cite{CampbellPRB1988,CampbellProc1989,CampbellPRB1990,AppelPRB1993,Hirsch1994,LuzPRB1996,SchillerPRB1999,Hubsch2006,Galvan2012}.

In our calculations $a$ plays a role of an effective CH parameter which is the only adjustable quantity, while all other parameters of the Hamiltonian (\ref{HamHubbard}) are taken from Ref. \cite{Chiesa2013}.
And we take into account five orbitals per ion, because $Cr$ ions in the molecular ring $Cr_8$ have five \emph{d-orbitals}. Their energies split due to the crystal field to form a lower energy quasi-triplet and a higher energy quasi-doublet. The energy splitting of the quasi-triplet is about $0.1eV$ and of the quasi-doublet about $0.05eV$, with the doublet energy being about $2eV$ higher than the triplet energy \cite{Chiesa2013}. Since there are three \emph{d-electrons} per $Cr$ ion, it is obvious that at the lowest temperatures they occupy orbitals belonging the quasi-triplet, whereas the states of the quasi-dublet are left unoccupied.

\section{Atomic limit: single ion results}

The Hilbert space dimension for 3 electrons occupying 3 orbitals is equal to 20.
Due to the intra-ion exchange couplings, the diagonalization of the single-ion part $H_0$ of the Hamiltonian (\ref{HamHubbard}) within this space results in formation of two quartets and six doublets. 
One of these quartets corresponding to the spin $S = 3/2$ is the ground state. 
Its energy $E_0$ is equal to 
\begin{equation}
E_0=\varepsilon_1+\varepsilon_2+\varepsilon_3+3U-9J, 
\label{enfor3el}
\end{equation}
where $\varepsilon_m\equiv t^{ii}_{mm}$ (here we used the same notation for energies of the orbitals $m=1,2,3$ as it is given in Ref.\cite{Chiesa2013}).
The eigenstates belonging to this quartet are displayed in Table \ref{singleEigenstates3el}, where
their representations are given in the graphically intuitive basis of the form $(orbital~1|orbital~2|orbital~3)$.
These states can be used to build up $4^N$ times degenerated N-ion ground state of N ions in the $t^{ii'}_{mn}=0$ limit. The full Hilbert space is then a tensor product of N spaces, each of which is spanned by the states given in Table \ref{singleEigenstates3el}, but corresponding to different ions.

\begin{table}[!htbp]
\caption{Eigenstates of $H_0$ forming the ground state quartet $S=3/2$ for a single-ion with 3 electrons and 3 orbitals.}
\label{singleEigenstates3el}
\centering
\resizebox{\columnwidth}{!}{%
\begin{tabular}{| l | c |}
\hline
$Sz$& State representation \\ \hline
$3/2$ &  $\qty(\uparrow | \uparrow | \uparrow)$  \\ \hline
$1/2$ &  $\dfrac{1}{\sqrt{3}} \qty\Big( \qty(\downarrow | \uparrow | \uparrow)+\qty(\uparrow | \downarrow | \uparrow)+\qty(\uparrow | \uparrow | \downarrow))$  \\ \hline
$-1/2$ &  $\dfrac{1}{\sqrt{3}} \qty\Big( \qty(\downarrow | \downarrow | \uparrow)+\qty(\downarrow | \uparrow | \downarrow)+\qty(\uparrow | \downarrow | \downarrow))$  \\ \hline
$-3/2$ &  $\qty(\downarrow | \downarrow | \downarrow)$ \\ \hline
\end{tabular}
}
\end{table}

\begin{table}[!htbp]
\caption{Exchange constants (in \emph{meV}) relating to the Heisenberg model: 
$\Gamma_{exp}$ - deduced from the experimental data, 
$\Gamma_{Ref. [16]}$ - reported in Ref. \cite{Chiesa2013}, 
$\Gamma$ - from the perturbation theory without CH ($a=0$) and with CH ($a=0.05$);
$\Gamma^*$ - from the exact diagonalization without CH ($a=0$) 
and with CH ($a=0.06$).
Results of our calculations are given in the last four columns.}
\label{exchangeConstant}
\centering
\resizebox{\columnwidth}{!}{%
\begin{tabular}{|c|c|c|c|c|c|}
\hline
$\Gamma_{exp}$ & $\Gamma_{Ref. [16]}$ & 
\multicolumn{2}{|c|}{$\Gamma$} & 
\multicolumn{2}{|c|}{$\Gamma^*$} \\ \cline{3-6} 
&  &  $a=0$ & $a=0.05$  & $a=0$ & $a=0.06$ \\
\hline
1.46 & 1.65 & 1.7 & 1.46 & 1.75 & 1.46 \\ 
\hline
\end{tabular}
}
\end{table}

\section{Interaction between two ions}
In the model (\ref{HamHubbard}), the interaction between ions results from electron jumps between orbitals belonging to adjacent ions. If amplitudes of these jumps are small enough with respect to $U,$ then we can use the perturbative calculus.
In the case of two ions the ground state of the unperturbed Hamiltonian $H_0$ is 16th fold degenerated
and its energy is equal to $2E_0=2(\varepsilon_1+\varepsilon_2+\varepsilon_3)+6U-18J$.
Electron jumps reduce this degeneration because an effective exchange interaction (kinetic exchange) between total spins $S=3/2$ of the ions is then formed.

In our case, electrons can jump either between quasi-triplet states or from a quasi-triplet to an empty quasi-doublet and back to the quasi-triplet. There are nine amplitudes $t^{12}_{mm'}$, $(m, m'=1,2,3)$ describing electron jumps from ion 2 to ion 1 and nine amplitudes $t^{21}_{mm'}$, describing jumps from ion 1 to ion 2.
So there are altogether 18 amplitudes describing electron hops between the quasi-triplet states, but due to the symmetry relationships $t^{ij}_{mm'}=t^{ji}_{m'm}$ only 9 of them are independent.
Since in the ground state the orbitals belonging to the quasi-triplets are single occupied, the electron hops are only allowed when their spins are anti-parallel and the CH effect is present then.
On the other hand, there are no CH effect for electron hops from the quasi-triplet to unoccupied quasi-doublet orbitals

Taking into account the electron jumps in the 2nd rank perturbative calculus applied to (\ref{HamHubbard})
in the limit of small $t^{ii'}_{mn}$ one gets an effective Heisenberg Hamiltonian of interacting spins $S=3/2$ with the antiferromagnetic super-exchange coupling $\Gamma^{ii'}_{SE}$.
If following the discussion given in \cite{Chiesa2013} we take into account also a direct ferromagnetic Coulomb exchange term $\Gamma^{ii'}_{CE}$ between the ions (not present in the Hamiltonian (\ref{HamHubbard}), but obtained independently via the constrained LDA), then the final form of the effective Hamiltonian $H_{eff}$ is as follows.

\begin{equation}
H_{eff}=\frac{1}{2}\sum_{i,i'} \Gamma^{ii'} S_{i} \cdot S_{i'}
\label{HamHeisenberg}
\end{equation}
where $\Gamma^{ii'}=\Gamma^{ii'}_{CE}+\Gamma^{ii'}_{SE}$ and the sum is over all pairs (not ordered) of adjacent magnetic ions. 

The super-exchange coupling 
$\Gamma^{ii'}_{SE}$ resulting from electrons kinetic can be expressed as the sum of the following two contributions
\begin{equation}
\Gamma^{ii'}_{SE}=\Gamma^{ii'}_0+\Delta\Gamma^{ii'},
\label{couplingHeisenbergParts}
\end{equation}
where the main part $\Gamma^{ii'}_0$ comes from jumps of electrons between single occupied states belonging to the quasi-triplets
\begin{equation}
\Gamma^{ii'}_0=\frac{2}{9}\sum^3_{n=1}\sum^3_{n'=1}
\dfrac{\abs{t^{ii'}_{nn'}}^2+\abs{t^{i'i}_{nn'}}^2 }{U+2J+\varepsilon_n-\varepsilon_{n'}} (1-a)^2
\label{couplingHeisenberg0}
\end{equation}
and the second part $\Delta\Gamma^{ii'}$ results from electron jumps between single occupied states 
belonging to the quasi-triplet and unoccupied states belonging to the quasi-doublet
\begin{equation}
\begin{aligned}
\Delta\Gamma^{ii'}=\frac{2}{9}&\sum^3_{n'=1}\sum^5_{n=4}
\dfrac{\abs{t^{ii'}_{nn'}}^2+\abs{t^{i'i}_{nn'}}^2 }{U+\varepsilon_n-\varepsilon_{n'}} \\
-\frac{2}{9}&\sum^3_{n'=1}\sum^5_{n=4}
\dfrac{\abs{t^{ii'}_{nn'}}^2+\abs{t^{i'i}_{nn'}}^2 }{U-3J+\varepsilon_n-\varepsilon_{n'}}.
\end{aligned}
\label{couplingHeisenbergD}
\end{equation}
The factor $(1-a)^2 $ associated with the effect of CH occurs only in the formula for 
$\Gamma^{ii'}_0$ because only in this case an electron hops to the orbital which is already occupied by another electron with an opposite spin. On the other hand, $\Delta \Gamma^{ii'}$ does not depend on the parameter $a$, because then the electron hops to an unoccupied orbital.
As a result, only $\Gamma^{ii'}_0$, but not $\Delta \Gamma^{ii'}$ is reduced due to the effect of CH. The simple forms of denominators in (\ref{couplingHeisenberg0}) and (\ref{couplingHeisenbergD}) result from the simple expression for energy (see Eq. \ref{enfor3el}) of the ground state multiplet displayed in Table \ref{singleEigenstates3el}.

Since we assume that the interactions between adjacent $Cr$ ions in $Cr_8$ are the same, from now we omit the upper indices $i, i'$ in the coupling constants $\Gamma$, $\Gamma_0$ and $\Delta\Gamma$. 
And if we do not take into account the CH effect by assuming that $a=0$ and put into the formula (\ref{couplingHeisenberg0}) the parameters taken from Table I in Ref. \cite{Chiesa2013}, then it turns out that $\Gamma_0$ is equal to $4.847meV$. 
Taking into account in the calculations the hoppings of electrons between states belonging to the quasi-triplet and quasi-doublet on adjacent ions results in the formation of a small ferromagnetic contribution of about $-0.334$meV (resulting from (\ref{couplingHeisenbergD})), which slightly reduces the $\Gamma$ value.
Another reduction in $\Gamma$ is due to the direct ferromagnetic exchange between ions (not included in the Hamiltonian (\ref{HamHubbard})), whose value estimated in Ref. \cite{Chiesa2013} amounts $\Gamma_{CE}= -0.34meV$.
After these reductions we get $\Gamma=4.17meV$, which  is much more than $1.46meV$, the value resulting from the experimental data.

Quite recently it turned out that the value of the electron hopping amplitude $t_ {11}^{ij}$ given in Ref. \cite{Chiesa2013} is overestimated and instead of $t_ {11}^{ij}= - 0.231 eV$, it should be $t_ {11}^{ij}= - 0.131 eV$ \cite{Santini}.
If we include in our calculations this new value of the parameter $t_ {11}^{ij}$, then we get $\Gamma= 1.7 meV$, which is still too much.
Therefore, we expect that for the correct description of this system, the effect of CH cannot be ignored.

There is no doubt that in real systems this effect occurs and that in our case it will lead to a decrease in the value of some $t^{ii'}_{mm'}$ amplitudes, and thus to a diminution in the $\Gamma$ value.
However, we are not aware if any calculations were made for $Cr_8$ that would allow us to estimate the size of the CH parameter $a$. Of course, such calculations would be appreciated, but as long as they are not available to us, we fit the value of $a$ in such a way as to obtain 
$\Gamma$ close to $1.46meV$. However, we suggest performing \emph{ab initio} calculations that would allow to verify if the value of the parameter $a$, that we got from fitting to the experimental data can be obtained by an \emph{ab initio} method.

Let us add that when we derive the value of the magnetic coupling constant not from the expressions (\ref{couplingHeisenbergParts})-(\ref{couplingHeisenbergD}) on $\Gamma_{SE}$ but directly from the diagonalization of the  multi-orbital HM Hamiltonian given in (\ref{HamHubbard}), then in order to meet the condition for $\Gamma^*=1.46meV$ we get a slighly higher value of $a\approx 0.06$. This is because the distances between energy levels determined by the diagonalization of the multorbital HM (\ref{HamHubbard}) are slightly larger than the distances obtained from the diagonalization of $H_ {eff}$ (\ref{HamHeisenberg}). 

\begin{figure}
\includegraphics[width=0.9\columnwidth]{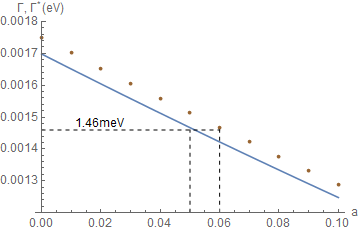}
\caption{The total exchange couplings: $\Gamma$ derived from the perturbation theory (the line) and $\Gamma^*$ derived from the exact diagonalization calculations (the dots) in $Cr_8$ as a function of the CH parameter $a$ for $U=5.98eV$, $J=0.26eV$ and the set of $t^{ii'}_{mm'}$ amplitudes reported in Ref. \cite{Chiesa2013} with the corrected value of $t_{11}^{ij}=-0.131 eV$ (colour online). The intersection points of the dashed lines indicate that the experimental value $\Gamma=1.46meV$ is attained either for $a\approx 0.05$ or $a\approx 0.06$, as obtained from the perturbation or exact diagonalization calculations, respectively.}
\label{Gamma-vs-a}
\end{figure}
\begin{figure}
\includegraphics[width=0.9\columnwidth]{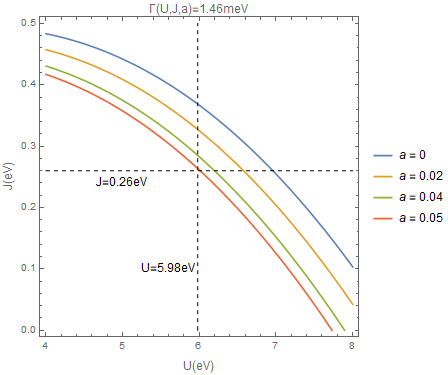}
\caption{The curves corresponding to the constant value of the superexchange coupling $\Gamma(U,J)=1.46meV$ in $Cr_8$ obtained from the perturbation calculus as a function of the interacting couplings $U$ and $J$ for the amplitudes $t^{ii'}_{mm'}$ reported in Ref. \cite{Chiesa2013}  with the corrected value of $t_{11}^{ij}=-0.131 eV$ and
for the following set of the CH parameters: $a= 0.0, 0.02, 0.04, 0.05$ (colour online). The intersection point of the dashed lines indicates that the experimental value $\Gamma=1.46meV$ is attained for $U=5.98eV$ and $J=0.26eV$ (the values reported in \cite{Chiesa2013}), when $a\approx 0.05$.}
\label{Jeffvarious-a}
\end{figure}

The procedure we used to determine $\Gamma^*$ by diagonalizing the multiorbital HM (\ref{HamHubbard}) was as follows. 
First we constructed the Hilbert spaces of states for the system of two ions with five orbitals per ion and fixed values of z-coordinate of the total spin $S_z$.
In our case, the dimensions of these spaces amounts 14400, 9450, 2520 and 210 for $S_z=0, \pm1, \pm2, \pm3$, respectively.

If there are no electron hopping between ions, then in all these ground-state multiplets (quartet for $S_z=0$, triplet for $S_z=\pm1$, doublet for $S_z=\pm2$ and singlet for $S_z=\pm3$) have the same energies. When the hopping is switched on, the multiplets split and for small hopping amplitudes the structure of the splitted multiplets is close to that occurring for the Heisenberg model.
Let us recall that for two spins $S=3/2$, that are coupled antiferromagnetically with the coupling constant $\Gamma$ the energy spectrum form four levels: $E=0, \Gamma, 3\Gamma$ and $6\Gamma$.

It is similar in the case of the multi-orbital HM when the hopping amplitudes are small enough. Indeed, for $S_z = 0$, the four lowest energy levels then have approximately the distribution in the form of $0$, $\Gamma^*$, $3\Gamma^*$, $6\Gamma^*$, but $\Gamma^*$ is slightly larger than $\Gamma$ ($\Gamma^*>\Gamma$).
Thus, the difference in energy between the lowest two levels can be assigned the coupling values $\Gamma^*$. In practice, it is more convenient to diagonalize HM not in the configuration space $S_z = 0$, but $S_z = 2$, because then we are dealing with the number of states equal to 2520 and not 14400. However, then the difference between the lowest two energy levels is approximately $3\Gamma^*$.

In Fig. \ref{Gamma-vs-a} we display an evolution of the total inter-ionic exchange couplings $\Gamma$ and $\Gamma^*$ with an increase of the CH coefficient $a$, when the on-site coupling parameters $U$ and $J$, as well as the amplitudes $t^{ii'}_{mm'}$ take the values reported in Ref. \cite{Chiesa2013} (with the corrected value of $t_{11}^{ij}=-0.131 eV$).
After viewing Fig. \ref{Gamma-vs-a} it is clear that $a \approx 0.05$ meets the condition $\Gamma=1.46meV$, whereas a slightly larger 
$a \approx 0.06$ meets the condition $\Gamma^*=1.46meV$. Indeed, we get: $\Gamma(a=0) \approx 1.7meV$ and 
$\Gamma^*(a=0) \approx 1.75meV$, whereas 
$\Gamma(a=0.05) \approx \Gamma^*(a=0.06) \approx 1.46meV$.

The impact of CH $g$ can be also noticed in Fig. \ref{Jeffvarious-a}, where the curves of constant value of $\Gamma=1.46meV$ for a few values of the parameter $a$ are displayed.
In this figure you can see that in order to get low enough $\Gamma$ for fixed $U$ and $J$, one should increase $a$ to the value of $a\approx 0.05$.

\section{Summary and conclusions}
In this contribution, we included the CH effect in the microscopic description of MNMs and showed that it reduces noticeably the value of the exchange constant between magnetic moments of the ions, thus bringing it closer to the value deduced from the experiments.
The final calculations we performed for $Cr_8$ using the data reported in \cite{Chiesa2013}.
By changing the parameter $a$ we found out that the agreement between the value of the exchange coupling $\Gamma$ calculated from the perturbation calculus and that deduced from the experiment is obtained for $a = 0.05$. A slightly higher value of $a\approx 0.06$ was needed when we performed our calculations by diagonalizing the Hamiltonian, but both these two values are quite small, which means that even a tiny CH effect is sufficient to reduce the exchange coupling to the value deduced from the experiment.
This  is why we believe that the explanation of the reason for the decrease in the antiferromagnetic coupling constant by means of CH, as proposed in this work, is justified.

There is no problem to extend these studies to other MNMs with various substitutions.
Indeed, both the model and the method used are universal and can be applied to other MNMs, but to obtain quantitative values for the magnetic exchange constants, microscopic parameters such as the local Coulomb coupling, the on-site exchange constant, the CH amplitudes and the correlated hopping parameters are needed from independent calculations.

\begin{acknowledgments}
The authors would like to thank G. Kamieniarz for useful discussion of some of the issues raised in this work and P. Santini for reading the manuscript and sharing his comments on it.
\end{acknowledgments}

\end{document}